# Addressing Trust Issues for Vehicle to Grid in Distributed Power Grids Using Blockchains

Yunwang Chen
*Department of Electrical and Electronic Engineering*
*Southern University of Science and Technology*
Shenzhen, China
chenyw2021@mail.sustech.edu.cn

Xiang Lei
*School of Mechanical and Electrical Engineering*
*Shenzhen Polytechnic University*
Shenzhen, China
leixiang@szpu.edu.cn

Linni Jian*
*Department of Electrical and Electronic Engineering*
*Southern University of Science and Technology*
Shenzhen, China
jianln@sustech.edu.cn
*Corresponding author

***Abstract*—While blockchain offers inherent security, trust issues among stakeholders in vehicle-to-grid (V2G) applications remain unresolved due to a lack of regulatory frameworks and standardization. Additionally, a tailored decentralized privacy-preserved coordination scheme for blockchain in V2G networks is needed to ensure user privacy and efficient energy transactions. This paper proposes a V2G trading and coordination scheme tailored to the decentralized nature of blockchain as well as the interests of stakeholders utilizing smart charging points (SCPs) and Stackelberg game model. Case studies using real-world data from Southern University of Science and Technology demonstrate the efficacy of the proposed scheme in reducing EV charging costs and the potential for supporting auxiliary grid services.**

***Keywords—electric vehicle, vehicle to grid, blockchain***

## I. INTRODUCTION

The rapid global adoption of electric vehicles (EVs) marks a significant shift towards sustainable transportation, promising substantial reductions in greenhouse gas emissions and decreased fossil fuel dependency [1]. According to [2], nearly 14 million electric cars were sold globally in 2023, marking a 35% increase from the previous year and bringing the total number of EVs on the road to 40 million. This growth trend is expected to continue, with electric car sales projected to reach around 17 million in 2024, representing more than one in five cars sold around the world. Governments worldwide have been enacting policies and providing incentives to promote the use of EVs, leading to a steady increase in their market penetration. This transition is driven not only by environmental concerns but also by advancements in EV technology, such as improved battery efficiency, longer driving ranges, and reduced production costs. As a result, EVs are becoming an increasingly viable alternative to traditional internal combustion engine vehicles, heralding a new era of cleaner and more efficient transportation.

Alongside the proliferation of EVs, the deployment of vehicle-to-grid (V2G) technology presents an innovative approach to energy management. V2G technology enables EVs to draw power from and supply electricity back to the grid, thus addressing the infrastructure pressure brought by the increasing number of EVs on the road [3]. This bidirectional energy flow has the potential to improve electricity management and distribution significantly. By providing additional resources during peak demand periods and absorbing excess power during times of low demand, V2G enhances grid stability and sustainability [4]. This dynamic interaction also facilitates the integration of renewable energy sources, such as wind and solar power, which are inherently variable and intermittent [5]. Acting as distributed energy storage units, EVs help balance supply and demand, reducing reliance on costly and polluting peaking power plants [6]. Forecasts indicate that by 2030, the energy storage capacity of electric vehicle batteries could sufficiently meet short-term grid storage requirements, underscoring the significant potential of EVs in enhancing grid stability and optimizing energy management [7].

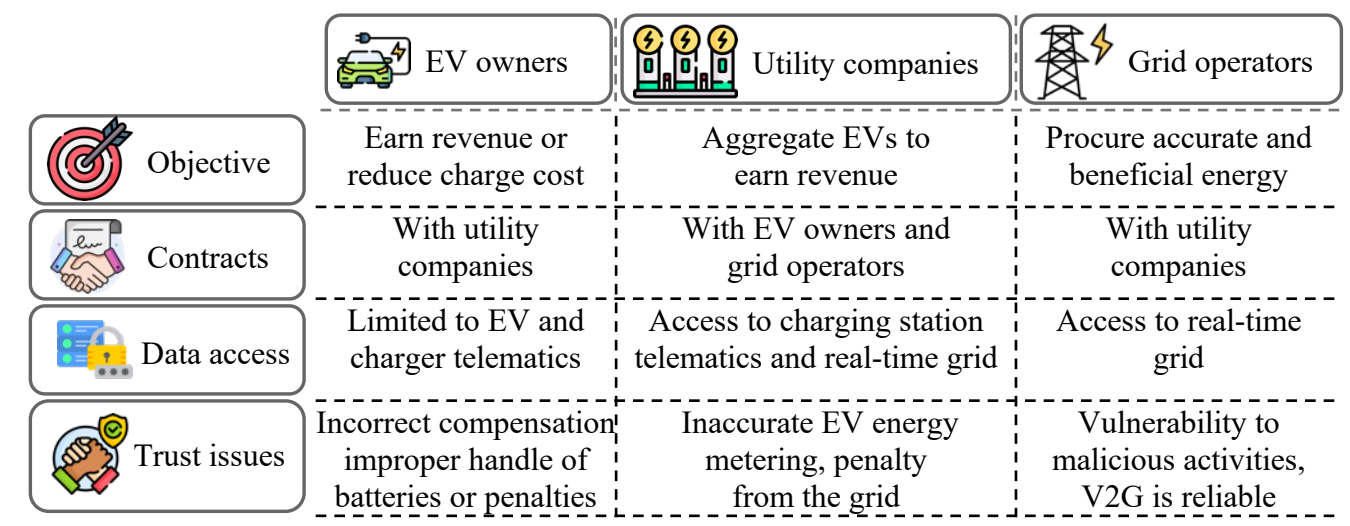

Fig. 1. An overview of trust issues for V2G in distributed power grid.

Despite the promising potential of V2G, several challenges impede its widespread implementation. One of the foremost issues is fostering trust among the diverse stakeholders involved, including EV owners, utility companies and grid operators [8]. Trust is crucial for successfully integrating V2G, as it involves complex interactions between various parties with different interests and priorities. As shown in Fig. 1, EV owners need to trust that their vehicles will be charged, receive correct compensation for providing V2G services, and be ready when needed. Meanwhile, utility companies and grid operators require assurance that energy monitoring of EVs is reliable [9]. Additionally, penalties based on the contract need to be backed by accurate monitoring and transactions. However, the interactions between EVs, utility companies, and the grid are comparatively vulnerable, making them attractive targets for malicious activities. Attackers may attempt to tamper with

This work was supported by the Science and Technology Innovation Committee of Shenzhen under Project JCYJ20220530113008019 and the Undergraduate Training Programs for Innovation and Entrepreneurship of Southern University of Science and Technology under Project 2024S14.

transactions or meters, such as modifying energy transfer amounts or times, which can disrupt the balance and efficiency of the grid. This disruption can lead to a loss of interest from EV users, utility companies, and grid operators [10].

Blockchain technology offers a decentralized and cryptographic framework that significantly mitigates the risks associated with cyber-attacks [11]. By utilizing a network of nodes to verify transactions, blockchain ensures that the consensus mechanism will detect and thwart most attempts to alter transaction data. Additionally, using public and private cryptographic keys enhances security, making it difficult for attackers to perform attacks without direct access to the user's private key. Blockchain's implementation of unique nonces and timestamps for each transaction further prevents replay attacks by ensuring that once a transaction is recorded, it cannot be reused or modified in ideal situations [12].

While blockchain is inherently secure, the trust issues among stakeholders (e.g., EV owners, utility companies, and grid operators) still need to be fully addressed. The need for more regulatory frameworks and standardization for blockchain in V2G applications poses a significant hurdle. Additionally, a tailored decentralized privacy-preserved coordination scheme for blockchain deployment in V2G networks still needs to be improved, which is essential to safeguard user privacy while ensuring efficient and transparent energy transactions. In this paper, we propose to design a decentralized V2G trading scheme tailored to blockchain features, aiming at improving transaction efficiency and scalability and streamlining the trading procedure. The main contributions of this paper are summarized as follows:

- A decentralized V2G trading framework is designed to utilize blockchain to improve transaction security and transparency. SCPs act as essential cyber-physical entities managing smart contracts between EVs and the power grid, facilitating real-time data exchange and control.
- A game-theoretical pricing strategy is proposed to optimize the trading process and maximize economic benefits. This strategy employs smart contracts for autonomous decision-making between EVs and operators, ensuring efficient and secure transactions. The dynamic pricing model adapts to market conditions, redistributing EV charging loads from peak to off-peak periods to enhance grid stability and reduce operational costs.
- The proposed framework is validated through a case study utilizing real-world data from the Southern University of Science and Technology (SUSTech). The case study demonstrates reductions in EV charging costs and highlights the framework's potential to support auxiliary grid services. The results affirm the effectiveness of the proposed V2G trading scheme.

The balance of the paper is organized as follows. Section II provides background that includes a review of V2G coordination and blockchain. Section III discusses the V2G trading framework and the pricing strategy. Subsequently, Section IV presents the results of case studies to validate the efficacy of the proposed scheme. Section V concludes the paper.

## II. Related Work

### A. V2G Coordination

V2G coordination relies heavily on effective scheduling, which can be centralized or decentralized [13]. The centralized framework allows the grid operator to directly control and coordinate the charging activities, mitigating the risks associated with uncoordinated charging, such as grid congestion and voltage fluctuations [14]. For instance, a centralized stochastic optimization model is employed in [15] to manage the spatiotemporal uncertainties of EVs and electricity markets, thus enhancing operational efficiency and economic benefits. In [16], a smart charging algorithm incorporating dynamic charging priority is proposed to manage EV charging loads, aiming to reduce computational complexity while introducing greater user flexibility. However, the practical application of centralized methods encounters hurdles due to the temporal and spatial variability of EV charging demands and the persisting computational complexity of deriving optimal solutions, which pose considerable challenges in real-world V2G scenarios [17]. Meanwhile, a decentralized V2G scheme is presented in [18], incorporating distributed computing capacity by engaging a network of interconnected smart charging points and dynamically responding to local conditions and demands. Besides, a decentralized V2G coordination method is proposed in [5], integrating the internet of smart charging points with photovoltaic systems and leveraging edge computing at each smart charging point to handle local data processing and decision-making, thereby reducing latency and enhancing privacy. These decentralized approaches offer promising solutions to the limitations of centralized methods, providing greater flexibility in V2G coordination.

### B. Energy Blockchain

With its security, transparency, and automation capabilities, blockchain technology emerges as a promising solution to enhance trust and efficiency in V2G systems by leveraging its inherent cyber-physical capabilities [19]. For instance, blockchain can facilitate secure and decentralized energy trading, allowing consumers to trade energy directly with each other, where blockchain technology promotes distributed or wholesale energy transactions and creates virtual grids [20]. Moreover, integrating blockchain with IoT devices for real-time monitoring and management of energy consumption is another promising application, as discussed in [21], offering a solution to address security and privacy challenges in the energy internet. Besides, a decentralized charging right transaction mechanism among green charging stations based on blockchain techniques is proposed in [22]. By incorporating distributed computing and blockchain, the system handles charging rights assignments through a complete P2P trading process, dynamically responding to local conditions and transaction risk factors. In conclusion, blockchain needs to meet the specific requirements of V2G systems for better performance.

## III. Methodology

### A. V2G System

The V2G system in the distributed power grid studied in this paper is depicted in Fig. 2. The system is primarily composed of the following entities:

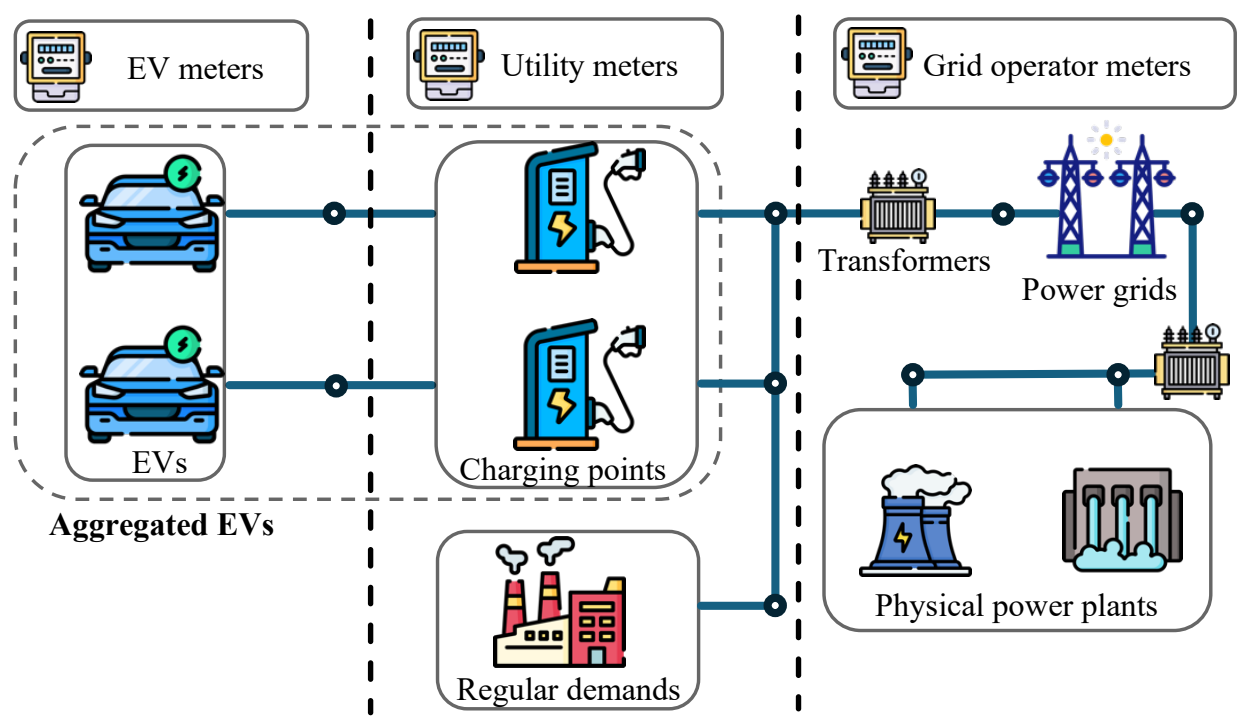

Fig. 2. The V2G systems in distributed power grid.

**EVs:** In V2G systems, EVs are considered prosumers, entities that produce and consume energy. Equipped with storage capacities, EVs can inject energy back into the grid or draw energy as needed. Their decision-making processes are influenced by economic incentives, adapting their energy dispatch in response to dynamic pricing and demand signals. Each EV is equipped with inherent metering capabilities that track energy consumption and injection.

**Utility company:** The utility company acts as an aggregator, intermediating between individual EV owners and the power grid. It aggregates the energy capacities of a fleet of EVs to buy or sell energy in bulk, enhancing the flexibility and reliability of energy management and generating profits. The aggregator manages contracts, redistributes payments to EV owners, and accrues profits through V2G administration fees. It also negotiates with the upper-level system operator to secure incentives for fulfilling auxiliary service obligations.

**Power grid operator:** The power grid operator oversees the integration of multiple generation sources and manages both unidirectional and bidirectional electricity flows, which facilitates contributions from energy prosumer entities, such as electric vehicle aggregators (EVA), while ensuring a consistent supply to regular consumers, including residential households and public facilities. The grid operator employs metering infrastructure to monitor energy flows and uses its computational capabilities to optimize grid operations. Furthermore, the power grid operator serves as the upper-level system operator, setting various prices for the energy market, including electricity production prices, electricity retail prices, auxiliary service prices, and other tariffs. These pricing mechanisms are vital for balancing supply and demand, incentivizing participation of both physical and virtual power plants, and ensuring the financial viability of energy transactions in the V2G systems.

## *B. Pricing Strategy*

Acknowledging the varying interests among EV owners, utility companies, and grid operators, we employ a Stackelberg game to model these interactions, focusing on the strategic interplay between EVs and utility companies. In this model, the utility company acts as the leader, setting prices and incentives, while EV owners respond as followers, adjusting their energy consumption and production based on the provided price signals.

**Follower model**: The utility function for an EV includes three main components: the satisfaction of charging or discharging, the cost associated with battery degradation, and the capital cost or gain through these activities. For EV $v_i$ in time slot $t$, let $SOC_i^t$ represent the SOC and $x_i^t$ denote the normalized quantity of electricity charged or discharged while $E_0$ acts as the standard quantity of the electricity. $E_i^t$ measures its current amount of electricity and $E_i^{capacity}$ measures the total electricity capacity for $v_i$. The utility function for $v_i$ can be modeled as:

$$u_i^t(x_i^t, p_c^t, p_d^t) = \begin{cases} S_i^t - p_c^t x_i^t, 0 < x_i^t \le 1, \\ U_{i,idle}, x_i^t = 0, \\ S_i^t - BDC_i^t + p_d^t x_i^t, -1 \le x_i^t < 0, \end{cases} \quad (1)$$

$$S_i^t(x_i^t) = \begin{cases} w_i^t \, log_2(1 + x_i^t), 0 \le x_i^t \le 1, \\ w_i^t(log_2(2 + x_i^t) - 1), -1 \le x_i^t < 0, \end{cases} \quad (2)$$

where $S_i^t$ represents the user satisfaction functions as detailed in (2), and $w_i^t = \frac{\beta_i}{SOC_i^t} > 0$ reflects the willingness to charge, with $\beta_i > 0$ being a predefined constant denoting the preference of user $i$. The willingness is inversely related to the remaining SOC, which indicates that when the $SOC_i^t$ is low, $v_i$ tends to have a higher willingness $w_i^t$ to charge [23]. $p_c^t > 0$ and $p_d^t < 0$ represent the charging and discharging price ruled by the utility company. $U_{i,idle}$ is the utility value set by the owner of $v_i$ in the idle scenario. $BDC_i^t = a_i x_i^t > 0$ represents the battery degradation cost in discharging.

Under varying charging and discharging prices, an EV can decide to charge, discharge, or remain idle, prioritizing its self-interest. Given the concavity of the utility function for these scenarios, the follower model problem can be easily solved, allowing the vehicle to act in a way that maximizes its utility.

**Leader model:** In the charging phase, the EVA charges EVs with dynamic, real-time electricity prices $p_{real-time}^t$ together with a specific service fee $W_{service}^t$. In the idle phase, the EVA incurs a nominal fee $p_{delay}^t < 0$ as the cost to pay for EVs' acceptance of this specific charging mode together with the delay. In the discharging phase, the EVA considers the auxiliary service demand $E_{limit}^t$ issued by the upper-level system operator, as well as the utility function for each vehicle $v_i$ . Subsequently, the EVA adjusts the discharging price $p_d^t$ to optimize its revenue generation. The revenue maximization problem of EVA, at time slot $t$ is formulated as follows:

$$\max_{p_d^t} R^t = R_{service}^t + R_{grid,V2G}^t - C_{V2G}^t - C_{limit}^t \quad (3)$$

$$R_{service}^t = \sum_{v_i \in \mathcal{V}_{charge}^t} W_{service}^t E_0 x_i^t, \quad (4)$$

$$R_{grid,V2G}^t = W_{grid}^t \sum_{v_i \in \mathcal{V}_{discharge}^t} E_0 |x_i^t|, \quad (5)$$

$$C_{V2G}^t = \sum_{v_i \in \mathcal{V}_{idle}^t} p_{delay}^t + \sum_{v_i \in \mathcal{V}_{discharge}^t} p_d^t E_0 x_i^t, \quad (6)$$

$$C_{limit}^t = |E_{limit}^t - \sum_{v_i \in \mathcal{V}_{discharge}^t} E_0 |x_i^t|| \cdot \delta_t, \quad (7)$$

subject to

$$0 \le SOC_i^t \le 1, \forall v_i \in \mathcal{V}^t, \forall t \in T, \quad (8)$$

$$x_i^t = arg \max_{x_i^t \in [-1,1]} u_i^t(x_i^t, p_c^t, p_d^t), \quad (9)$$

$$(SOC_i^t - SOC_i^{t-1}) \cdot E_i^{capacity} = E_0 \cdot x_i^t, \\ \forall v_i \in \mathcal{V}^t, \forall t \in T, \quad (10)$$

$$p_{d,min}^t \le p_d^t \le p_{d,max}^t \le 0 \le p_{delay}^t, \forall t \in T, \quad (11)$$

where $\mathcal{V}_{charge}^t$, $\mathcal{V}_{discharge}^t$ and $\mathcal{V}_{idle}^t$ denote the set of EV charging, discharging and remaining idle at time slot $t$, respectively. In equation (4), $R_{service}^t$ indicates the service fee collected for the charging service provided by EVA at time slot $t$ and $W_{service}^t$ is the service fee for providing charging service. In equation (5), $R_{grid,V2G}^t$ signifies the revenue gained at the auxiliary market during time slot $t$, in which $W_{grid}^t$ is the auxiliary service price offered by the upper-level system operator. In equation (6), $C_{V2G}^t$ represents the cost incurred by the aggregator for V2G services at time slot $t$. In equation (7), $C_{limit}^t$ denotes the penalty for not fulfilling the discharging task issued by upper-level systems, where $E_{limit}^t \ge 0$ is the auxiliary service amount and $\delta_t \ge 0$ is the penalty factor for not achieving the demand. Equation (8) indicates the constraint of $SOC_i^t$ for EV $v_i$ attended in V2G activities. Equation (9) indicates that $v_i$ will decide its actions, i.e. charge, discharge, idle to obtain its best utility at time slot $t$. Equation (10) denotes the relationship between $SOC_i^t$ and energy trading amount, in which $E_i^{capacity}$represents the energy capacity of $v_i$. Equation (11) imposes regulation limitations on the discharging price [3].

### C. Decentralized V2G Trading Scheme

In order to design a decentralized V2G trading scheme tailored to blockchain features, we consider SCPs to be key components in proposed decentralized V2G systems for their monitoring, communication, storage, computation and control capabilities. Based on these capabilities, the SCPs are able to function as nodes in blockchain networks, with each EV $v_i, i = 1,2,3 \dots N$ is equipped with a digital wallet. $v_i^{plugged}$ is used to denote EV $v_i$, that have been connecting to a SCP, in which $v_i^{plugged}$ functions as a light client. This setup facilitates blockchain transactions without necessitating the storage of the entire blockchain [24]. Consequently, EVs are allowed to engage in blockchain transactions efficiently, verifying only essential information such as block headers without the necessity to store the entire blockchain. Each EV is endowed with a unique digital signature, ensuring authenticated transactions and preventing repudiation. SCPs, functioning as full nodes, maintain complete copies of the blockchain, perform parallel computations, validate transactions and blocks, and conduct operational trading on behalf of the $v_i$that the SCP is physically linked with. Oracles bridge the system with external world, providing secure, reliable, and timely data inputs, such as real-time electricity price $p_{real-time}^t$ in the retail market [11].

The smart contract process is articulated through the following steps:

*1) Initialization:* An EV $v_i$ is plugged into the SCP $SCP_i$ with a preset preference $\beta_i$ and an idle utility choice $U_{i,idle}$ determined by the EV owner. $SCP_i$ verifies the identity of the EV owner and the plugged-in EV $v_i^{plugged}$.

*2) Information transfer:* Each $v_i^{plugged}$ conveys the requisite information to its corresponding $SCP_i^{plugged}$, signed by the digital signature of the wallet owner. The $SCP_i^{plugged}$ validates the status of $v_i^{plugged}$ and saves the parameters for the follower model to the local ledger. Subsequently, $SCP_i^{plugged}$ disseminates relevant data through its communication network.

*3) Data assimilation:* The system reads the current time slot $t$ and assimilates essential data such as the real-time charging price $p_t^{real\text{-}time}$, the permissible range of discharging prices $[p_{d,min}^t, p_{d,max}^t]$ for the current time slot, the auxiliary service task $E_t^{limit}$, and the auxiliary service price $W_t^{service}$ as communicated by the upper-level system operator via oracles.

*4) Discahrging price calculation:* The EVA calculates the optimal discharging price with precision $\epsilon$ :

$$p_d^{t\star} = ar\,g \max_{p_d^t \in [p_{d,min}^t, p_{d,max}^t]} R^t(p_d^t). \quad (12)$$

This computation is distributed among available functional SCPs, each handling a subrange of $p_d$ to reduce the complexity of the leader model. The subranges are dynamically allocated based on the computational capacity of each SCP, ensuring efficient resource utilization.

*5) Solution logging and network broadcasting:* The solutions are recorded in the ledger and voted on by the nodes. Once the optimal $p_d^*$ is determined and verified, it is broadcast across the network. Each $SCP_i^{plugged}$ then calculates follower model$x_i^t = \arg\max_{x_i \in [0,1]} u_i^t(x_i, p_t^c, p_d^*)$ to respond to the current real-time charging price and the established discharging price $p_d^*$. Following this, $SCP_i$ conducts the power transmission $x_i^t$ with the connected $v_i^{plugged}$, and $v_i^{plugged}$ pays the corresponding fees from its wallet.

*6) Completion and preparation for the next cycle:* Upon completion of the operational cycle, the system enters a preparatory phase for the upcoming time slot. Here, each $v_i^{plugged}$, as a light client, updates its internal blockchain ledger with the latest transaction information by accessing the connected $SCP_i^{plugged}$ and stores it in the EV's internal memory, ready for the next operational cycle starting at step 2 [25].

## IV. Case Study

The case studies utilize historical EV routines obtained from SUSTech within the Guangdong electricity market in China. The main parameter settings are detailed in TABLE I. This study examines a dataset size of 2,000 EVs, comprising electric buses and privately owned EVs. The dataset includes EV parameters such as battery capacity (in kWh), initial battery state of charge (in kWh), and EV arrival and departure times (in daytime), as illustrated in Table II with six representative samples. The real-time charging electricity pricing model follows the valley-peak pricing scheme implemented by the Shenzhen government, which features three distinct pricing tiers: peak, valley, and flat rates, each applicable during specific periods of the day. The allocation of these rates is detailed in Table III [26].

TABLE I. MAIN PARAMETERS IN THE CASE STUDIES

| ***Parameter*** | ***Value*** | ***Parameter*** | ***Value*** |
|---|---|---|---|
| $W_{grid}^{t}$ | 0.792CNY/kWh | $W_{service}^{t}$ | 0.5CNY/kWh |
| $p_{d,max}^{t}$ | 0 | $p_{d,min}^{t}$ | $-3 \cdot p_{real-time}^{t}$ |
| $E_0$ | 3.75kWh | $U_{i,idle}$ | 0 |
| $p_{delay}^{t}$ | -0.1 CNY/15min | $\beta_i$ | 0.16 |
| $a_i$ | -0.01 CNY/kWh | $\delta_t$ | $2 \cdot p_{real-time}^{t}$ |

TABLE II. DRIVING PATTERNS SAMPLE OF EVS

| EV sample IDs | ***Battery volume (kWh)*** | ***Initial battery (kWh)*** | ***Arrival time (h)*** | ***Departure time (h)*** |
|---|---|---|---|---|
| 1 | 160 | 57.2 | 16:00 | next day 06:30 |
| 71 | 64 | 19.9 | 17:45 | next day 16:30 |
| 74 | 65 | 17.7 | 14:30 | next day 04:30 |
| 215 | 40 | 26.2 | 5:45 | 10:15 |
| 217 | 65 | 37.1 | 0:00 | 12:00 |
| 486 | 32 | 20.2 | 21:30 | next day 12:00 |

TABLE III. VALLEY-PEAK PRICES (CNY/KWH) AT EACH HOUR

| Time period | ***00:00-08:00*** | ***08:00-10:00*** | ***10:00-12:00*** | ***12:00-14:00*** | ***14:00-19:00*** | ***19:00-24:00*** |
|---|---|---|---|---|---|---|
| Type | Valley | Normal | Peak | Normal | Peak | Normal |
| Price | 0.26 | 0.66 | 1.12 | 0.66 | 1.12 | 0.66 |

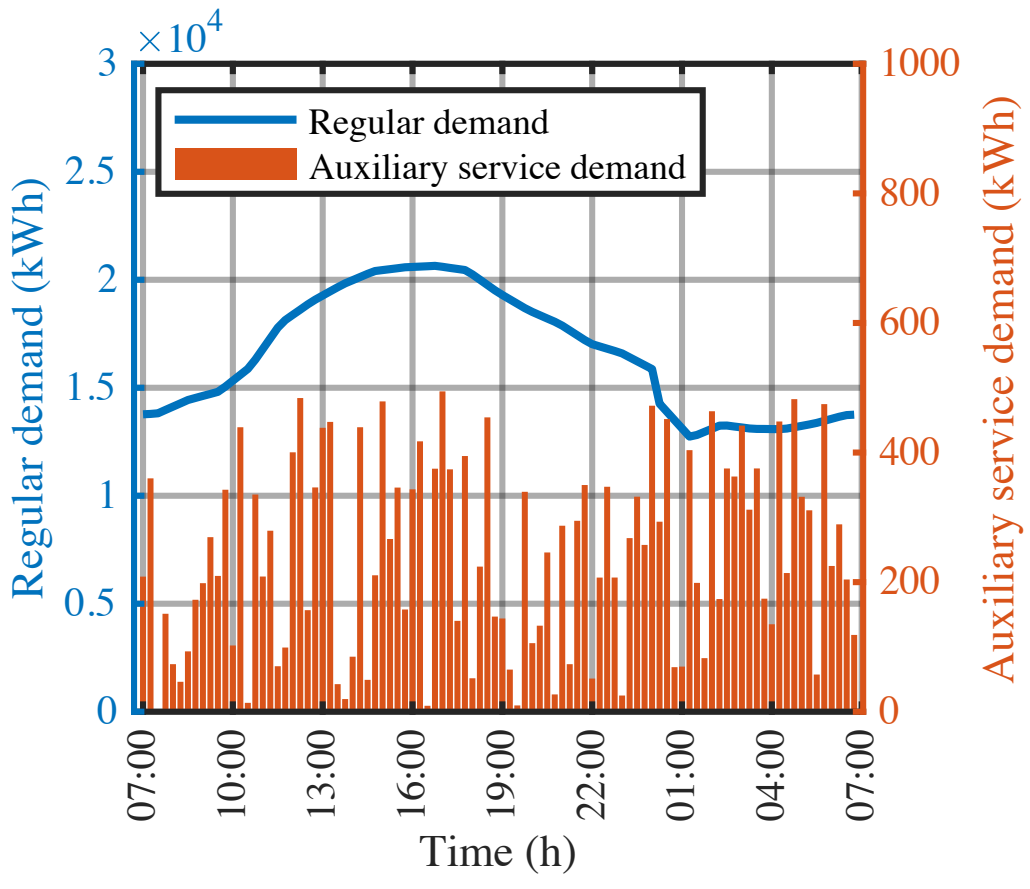

Fig. 3. Base load and simulated auxiliary demand in SUSTech campus from May 27 07:00 to May 28 07:00 2021.

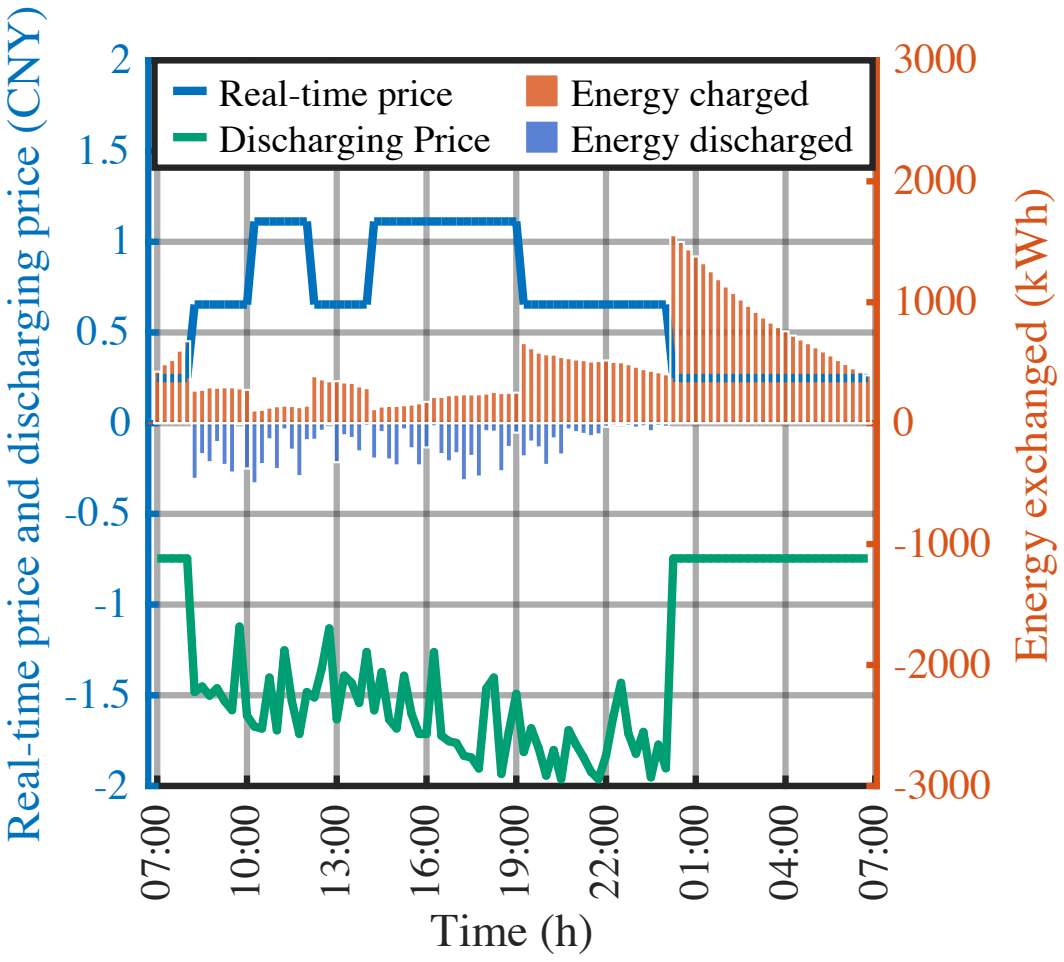

Fig. 4. Charging/discharging power with real-time electricity price and discharging price under the V2G scheme.

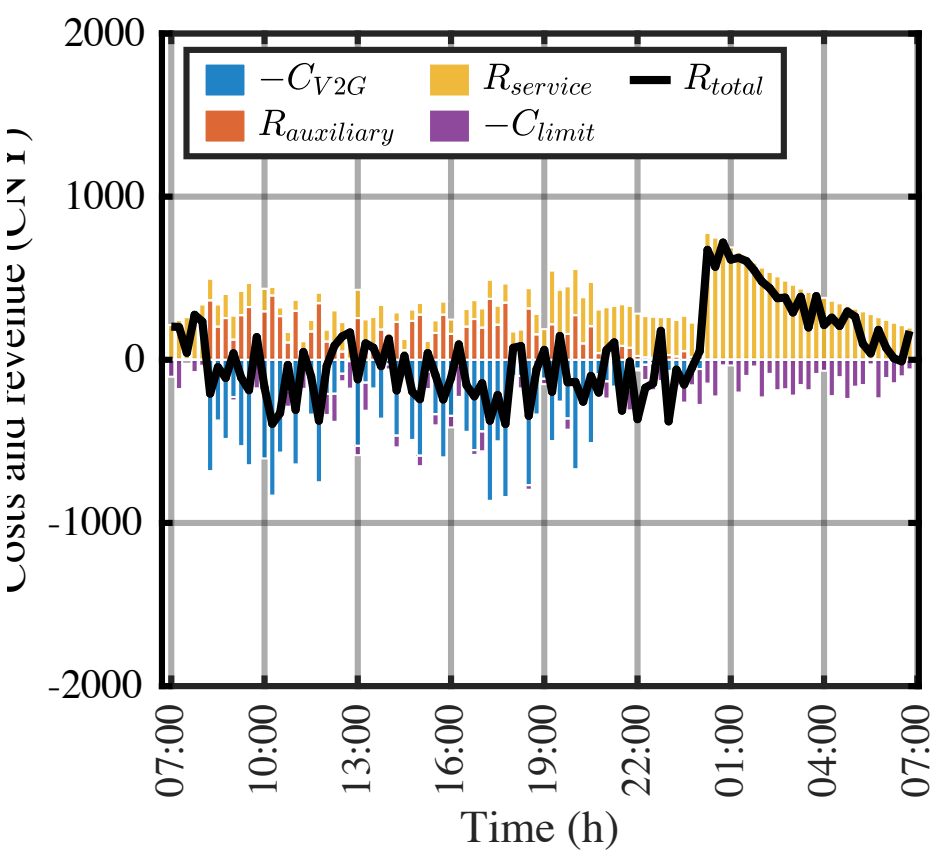

Fig. 5. Revenue and cost analysis of the V2G scheme.

Fig. 3 displays the actual loads for the SUSTech campus from 07:00 on May 27 to 07:00 on May 28, 2021, along with the simulated auxiliary service demand. Fig. 4 illustrates the bidirectional power flow from aggregated EVs participating in the proposed V2G scheme, real-time valley-peak electricity pricing in Shenzhen, and the calculated optimal discharging price. The correlation between the discharging price and the real-time electricity price demonstrates EVA's ability to adjust prices responsively to market conditions. The amount of energy exchange demonstrates SCPs' capability to schedule and trade on behalf of plugged-in EVs' interests. Finally, Fig. 5 illustrates the financial dynamics, including costs and revenues, associated with the proposed V2G scheme, utilizing notations introduced in Section III. From the EVA's perspective, the scheme enables efficient utilization of EVs to meet auxiliary service demand, thereby generating profits and reducing penalties. However, during periods of lowest electricity prices (the 'valley' stage), the EVA's ability to capitalize on the auxiliary services market becomes lower.

## V. CONCLUSION

This study examines the blockchain-enabled V2G trading scheme in the context of a distributed power grid. A decentralized V2G trading scheme is proposed, leveraging blockchain technology to enhance security, transparency, and efficiency in energy transactions, thereby fostering trust among stakeholders. By utilizing smart charging points (SCPs) as intermediaries, the proposed framework allows EVs to engage in secure information transfer and submit their energy demands without requiring extensive computational and storage capabilities. A Stackelberg game-based pricing strategy is proposed tailored to the decentralized nature of blockchain as well as the interests of EVA and EVs. The case study using real-world data from the Southern University of Science and Technology validates the effectiveness of the proposed scheme.

## ACKNOWLEDGMENT

This work was supported by the Science and Technology Innovation Committee of Shenzhen under Project JCYJ20220530113008019 and the Undergraduate Training Programs for Innovation and Entrepreneurship of Southern University of Science and Technology under Project 2024S14.